\documentclass{pasj00}
\begin{document}
\SetRunningHead{T. Kato et al.}{Large-Amplitude Oscillation in V425 Cas}

\Received{}%{yyyy/mm/dd}
\Accepted{}%{yyyy/mm/dd}

\title{Large-Amplitude 2.65-d Oscillation in the VY Scl-Type star V425 Cas}

\author{Taichi \textsc{Kato}, Makoto \textsc{Uemura}, Ryoko \textsc{Ishioka}}
\affil{Department of Astronomy, Faculty of Science, Kyoto University,
       Sakyou-ku, Kyoto 606-8502}
\email{tkato@kusastro.kyoto-u.ac.jp, uemura@kusastro.kyoto-u.ac.jp,
       ishioka@kusastro.kyoto-u.ac.jp}

\author{Timo \textsc{Kinnunen}}
\affil{Sinirinnantie 16, SF-02660 Espoo, Finland}
\email{stars@personal.eunet.fi}

%%% end:list of authors

\KeyWords{Accretion: accretion disks
          --- Stars: cataclysmic variables
          --- Stars: dwarf novae
          --- Stars: oscillations
          --- Stars: individual (V425 Cas)}

\maketitle

\begin{abstract}
  From long-term photometry of a VY Scl-type star, V425 Cas, between
1998 and 2000, we discovered a short-term, large-amplitude (up to 1.5
mag) variations.  The variation was well represented by a single period
of 2.65 d.  The large amplitude and the profile of the folded light
curve suggest that the dwarf nova-type disk instability is responsible
for this variation.  The shortness of the period is unprecedented in
hydrogen-rich cataclysmic variables.  Given the recent emerging
evidence that the irradiation from white dwarfs in VY Scl-type systems
affect their light behavior, we propose a possibility that this unique
variation in V425 Cas can be explained by the combination of the dwarf
nova-type disk instability and irradiation.  Similar short-period
``outbursts" have been known in X-ray transients (V518 Per), and helium
cataclysmic variables (CR Boo and V803 Cen).  We discuss the possibility
that these phenomena have a common origin to the unique variation in
V425 Cas.
\end{abstract}

\section{Introduction}
   Cataclysmic variables (CVs) are close binary systems consisting of
a white dwarf and a red dwarf secondary filling the Roche lobe.
The matter transferred from the secondary forms an accretion disk
around the white dwarf.  Instabilities in the accretion disk result
in various kinds of activities seen in CVs.  The two most relevant
instabilities are thermal and tidal instabilities, which are responsible
for dwarf nova-type outbursts and superhumps, respectively (see
\citet{osa96} for a review).  In systems having orbital periods longer
than three hours, thermal instability governs the general behavior.
Accretion disks become thermally stable at high mass-transfer rates
($\dot{M}$), while disks becomes thermally unstable in low $\dot{M}$.
The former corresponds to novalike (NL) variables, which do not show
outbursts, while the latter corresponds to dwarf novae, which show
semi-periodic outbursts.  Close to the thermal stability border, there
exist systems called as Z Cam stars, which show both standstills,
the state corresponding to a NL variable, and a state exhibiting dwarf
nova outbursts.  The origin of interchanging states in Z Cam stars
can be naturally understood with an assumption of varying $\dot{M}$
from the secondary: the high $\dot{M}$-state corresponds to standstills,
while the low $\dot{M}$-state the dwarf nova phase.  There exist
a small group of NL variables which show temporary reduction or
cessation of $\dot{M}$ from the secondary, namely VY Scl-type stars
\citep{war95}.  The low states (reduced $\dot{M}$-states) of VY Scl
stars are hard to interpret in the scheme of the standard disk
instability theory.  If the disk follows the same evolution as in
Z Cam stars in response to the temporary reduction of $\dot{M}$,
the system should undergo dwarf nova outbursts (\cite{hon94},
\cite{kin98}).  Observations usually show the contrary: the system
undergoes a smooth decline from the high to low states (\cite{hon94};
see also \cite{gre98}).

   An important clue to understanding the VY Scl-type behavior came from
the detection of transient supersoft X-rays from a VY Scl-type system,
V751 Cyg \citep{gre99}.  From the detection of supersoft X-rays during
the low state of V751 Cyg, \citet{gre99} suggested that steady nuclear
burning is taking place on the white dwarf of V751 Cyg.  \citet{gre99}
further suggested the possibility that VY Scl-type stars comprise
a low-mass analog of supersoft X-ray sources (SSXS).  This discovery
is consistent with the observed high temperature of white dwarfs
in VY Scl-type systems (cf. \cite{war95}, table 2.8).  \citet{lea99}
proposed that, in the presence of the heating from the hot white
dwarf, the irradiation on the accretion disk suppresses the thermal
instability, which can reproduce the observed light curve of VY Scl-type
system in their low-high transitions.  This effect, combined with
the suggestion by \citet{gre99}, would be a promising candidate for
the explanation of the behavior of VY Scl-type stars.  In this paper,
we report on the discovery of large-amplitude oscillations with a period
of 2.65 d, in a VY Scl-type system V425 Cas \citep{wen87}, which we
regard as a further evidence for the effect of irradiation.

\section{Observation}
   A fading of V425 Cas was detected by one of authors (T. Kinnunen)
in 1997 August.  The object soon returned to its high state, and remained
at around mag 14.5 until early 1998.  The object was again caught in faint
state at the beginning of the next observing season (1998 August).
We observed the system with a CCD in three seasons, 1998 -- early 1999,
late 1999 -- early 2000 and early 2000.  The CCD observations were done
using an unfiltered ST-7 camera attached to the Meade 25-cm
Schmidt-Cassegrain telescope.  The exposure time was 30 s.  The images were
dark-subtracted, flat-fielded, and analyzed using the Java$^{\rm TM}$-based
PSF photometry package developed by one of the authors (T. Kato).
The magnitudes were determined relative to GSC 3985.1444 (Tycho-2 magnitude:
$V=11.07\pm0.17, B-V=+0.17\pm0.10$), whose constancy was confirmed using
GSC 3985.1525.  The log of observations are given in the table ($N$
represents the number of frames).

\begin{table}
\caption{Log of observations (1)}\label{tab:table1}
\begin{center}
\begin{tabular}{ccccc}
\hline
Month    & UT (start--end)  &  N  & mag   & error \\
\hline
1998     &                  &     &       &       \\
December & 12.454 -- 12.505 & 131 & 4.625 & 0.148 \\
         & 13.369 -- 13.494 & 304 & 5.224 & 0.241 \\
         & 15.366 -- 15.491 & 252 & 5.048 & 0.831 \\
         & 16.368 -- 16.493 & 303 & 5.129 & 0.167 \\
         & 17.393 -- 17.469 &  69 & 5.517 & 1.860 \\
         & 18.398 -- 18.487 & 208 & 4.945 & 0.155 \\
         & 19.366 -- 19.495 & 308 & 4.559 & 0.200 \\
         & 20.381 -- 20.465 & 185 & 4.366 & 0.208 \\
         & 21.365 -- 21.446 & 202 & 4.309 & 0.130 \\
         & 22.360 -- 22.409 &  36 & 5.298 & 1.241 \\
         & 23.354 -- 23.436 & 199 & 4.437 & 0.166 \\
         & 25.359 -- 25.468 & 267 & 5.196 & 0.185 \\
         & 26.365 -- 26.501 & 175 & 5.201 & 1.084 \\
         & 27.364 -- 27.422 & 139 & 4.756 & 0.112 \\
         & 28.414 -- 28.444 &  78 & 4.438 & 0.129 \\
         & 29.358 -- 29.397 & 100 & 5.452 & 0.397 \\
         & 30.358 -- 30.420 & 160 & 4.498 & 0.248 \\
         & 31.374 -- 31.381 &   5 & 3.865 & 0.593 \\
1999     &                  &     &       &       \\
January  &  1.354 --  1.362 &  19 & 5.307 & 0.317 \\
         &  2.367 --  2.371 &   9 & 4.230 & 0.246 \\
         &  3.355 --  3.366 &  28 & 4.988 & 0.606 \\
         &  4.383 --  4.391 &  20 & 5.198 & 0.265 \\
         &  5.358 --  5.370 &  31 & 4.869 & 0.175 \\
         &  7.360 --  7.365 &  14 & 4.927 & 0.606 \\
         &  8.359 --  8.367 &  19 & 4.353 & 0.175 \\
         &  9.365 --  9.375 &  26 & 5.197 & 1.102 \\
         & 10.362 -- 10.368 &  14 & 4.610 & 0.204 \\
         & 11.365 -- 11.369 &  10 & 4.350 & 0.125 \\
         & 12.363 -- 12.371 &  20 & 5.669 & 0.594 \\
         & 13.363 -- 13.368 &  11 & 4.553 & 0.318 \\
         & 14.433           &   1 & 3.819 &  --   \\
         & 15.364 -- 15.371 &  19 & 5.401 & 0.643 \\
\hline
\end{tabular}
\end{center}
\end{table}

\begin{table}
\caption{Log of observations (2)}\label{tab:table2}
\begin{center}
\begin{tabular}{ccccc}
\hline
Month    & UT (start--end)  &  N  & mag   & error \\
\hline
1999     &                  &     &       &       \\
January  & 16.377 -- 16.384 &   8 & 4.123 & 0.527 \\
         & 17.388 -- 17.393 &  14 & 5.344 & 0.217 \\
         & 18.366 -- 18.372 &  16 & 5.104 & 0.359 \\
         & 20.421 -- 20.426 &  14 & 4.294 & 0.110 \\
         & 21.367 -- 21.371 &  10 & 4.060 & 0.613 \\
         & 22.390 -- 22.394 &  12 & 5.051 & 0.178 \\
         & 23.368 -- 23.375 &  20 & 5.069 & 0.651 \\
         & 25.418 -- 25.424 &  15 & 4.565 & 0.986 \\
         & 26.371 -- 26.382 &  28 & 4.380 & 0.164 \\
         & 27.536 -- 27.540 &  12 & 4.663 & 0.315 \\
         & 28.377 -- 28.380 &  11 & 5.187 & 0.347 \\
         & 29.374 -- 29.379 &  13 & 4.299 & 0.150 \\
         & 30.373 -- 30.379 &  16 & 5.101 & 0.379 \\
         & 31.375 -- 31.379 &   9 & 4.918 & 0.236 \\
February &  1.427 --  1.445 &  49 & 4.771 & 0.401 \\
         &  2.375 --  2.384 &  23 & 4.825 & 0.442 \\
         &  4.564 --  4.574 &  23 & 4.848 & 1.164 \\
         &  5.434 --  5.439 &   4 & $>$4.5&  --   \\
         &  6.378 --  6.386 &  21 & 5.109 & 0.326 \\
         &  7.386 --  7.401 &  39 & 4.456 & 0.148 \\
         &  8.379 --  8.398 &  49 & 5.320 & 0.422 \\
         &  9.383 --  9.396 &  33 & 4.530 & 0.512 \\
         & 10.385 -- 10.396 &  15 & 4.756 & 1.207 \\
         & 13.410 -- 13.427 &  19 & 4.807 & 0.749 \\
         & 14.385 -- 14.395 &  26 & 5.282 & 0.373 \\
         & 15.386 -- 15.410 &  38 & 4.470 & 0.931 \\
         & 16.386 -- 16.392 &  18 & 5.411 & 0.926 \\
         & 17.393 -- 17.397 &  10 & 4.241 & 0.241 \\
         & 19.404 -- 19.406 &   5 & 4.159 & 0.086 \\
         & 20.390 -- 20.396 &  15 & 5.174 & 0.507 \\
         & 21.399 -- 21.401 &   7 & 4.869 & 0.511 \\
         & 22.390 -- 22.394 &   9 & 4.400 & 0.381 \\
         & 23.425 -- 23.427 &   6 & 4.860 & 0.286 \\
\hline
\end{tabular}
\end{center}
\end{table}

\begin{table}
\caption{Log of observations (3)}\label{tab:table3}
\begin{center}
\begin{tabular}{ccccc}
\hline
Month    & UT (start--end)  &  N  & mag   & error \\
\hline
1999     &                  &     &       &       \\
February & 25.394 -- 25.396 &   6 & 4.257 & 0.284 \\
         & 27.408 -- 27.412 &   4 & 3.850 & 0.495 \\
         & 28.393 -- 28.399 &  16 & 4.277 & 0.214 \\
March    &  1.397 --  1.403 &  15 & 5.013 & 0.369 \\
         &  3.396 --  3.401 &  11 & 5.246 & 0.689 \\
         &  5.402 --  5.408 &  10 & 5.147 & 1.049 \\
         &  6.397 --  6.401 &  12 & 4.878 & 0.876 \\
         & 10.399 -- 10.407 &  21 & 4.915 & 0.859 \\
         & 11.400 -- 11.407 &  14 & 5.027 & 1.206 \\
         & 16.404 -- 16.411 &  19 & 6.527 & 2.544 \\
October  &  1.669 --  1.672 &  10 & 4.920 & 0.141 \\
         &  3.668           &   1 & 5.352 &  --   \\
         &  5.636 --  5.639 &   8 & 5.185 & 0.167 \\
         &  9.710 --  9.712 &   2 & 4.952 & 0.210 \\
         & 10.646 -- 10.674 &  67 & 4.773 & 0.322 \\
         & 11.647 -- 11.649 &   5 & 4.840 & 0.109 \\
         & 17.636 -- 17.640 &  10 & 4.874 & 0.201 \\
         & 18.651 -- 18.651 &   2 & 4.565 & 0.501 \\
         & 20.603 -- 20.606 &   8 & 4.786 & 0.106 \\
         & 22.613 -- 22.616 &  10 & 4.705 & 0.198 \\
         & 23.611 -- 23.615 &  10 & 4.758 & 0.148 \\
         & 24.610 -- 24.613 &  10 & 4.820 & 0.097 \\
         & 25.601 -- 25.605 &  10 & 4.779 & 0.127 \\
         & 28.646 -- 28.650 &  10 & 4.940 & 0.590 \\
         & 29.588 -- 29.594 &  17 & 4.754 & 0.303 \\
         & 30.577 -- 30.581 &  11 & 4.604 & 0.327 \\
November &  3.562 --  3.563 &   5 & 4.654 & 0.070 \\
         &  4.562 --  4.564 &   5 & 4.661 & 0.086 \\
         &  5.554 --  5.557 &   8 & 4.671 & 0.140 \\
         & 19.565 -- 19.567 &   5 & 4.588 & 0.062 \\
         & 21.562 -- 21.564 &   5 & 4.791 & 0.140 \\
         & 22.564 -- 22.565 &   5 & 4.669 & 0.271 \\
         & 25.753 -- 25.757 &  11 & 4.659 & 0.612 \\
\hline
\end{tabular}
\end{center}
\end{table}

\begin{table}
\caption{Log of observations (4)}\label{tab:table4}
\begin{center}
\begin{tabular}{ccccc}
\hline
Month    & UT (start--end)  &  N  & mag   & error \\
\hline
1999     &                  &     &       &       \\
November & 26.754 -- 26.757 &  10 & 4.877 & 0.266 \\
         & 28.688 -- 28.692 &  11 & 4.913 & 0.171 \\
         & 29.543 -- 29.544 &   5 & 4.649 & 0.148 \\
         & 30.543 -- 30.544 &   3 & 4.783 & 0.137 \\
December &  3.656 --  3.659 &  10 & 4.649 & 0.131 \\
         &  7.652 --  7.656 &  10 & 4.698 & 0.183 \\
         &  8.637 --  8.640 &  10 & 4.879 & 0.204 \\
         &  9.478 --  9.479 &   5 & 4.699 & 0.182 \\
         & 10.473 -- 10.474 &   5 & 4.469 & 0.151 \\
         & 11.463 -- 11.464 &   5 & 4.877 & 0.486 \\
         & 14.405 -- 14.406 &   5 & 4.843 & 0.099 \\
         & 15.407 -- 15.409 &   5 & 4.720 & 0.100 \\
         & 16.413 -- 16.414 &   5 & 4.843 & 0.068 \\
         & 19.424 -- 19.426 &   5 & 4.585 & 0.253 \\
         & 20.419 -- 20.420 &   5 & 4.485 & 0.199 \\
         & 21.419 -- 21.420 &   5 & 4.797 & 0.122 \\
         & 22.422 -- 22.423 &   5 & 4.632 & 0.170 \\
         & 23.412 -- 23.413 &   5 & 4.902 & 0.104 \\
         & 24.412 -- 24.413 &   4 & 4.736 & 0.196 \\
         & 25.413 -- 25.415 &   5 & 4.468 & 0.097 \\
         & 26.412           &   1 & 4.356 &  --   \\
         & 27.405 -- 27.406 &   5 & 4.721 & 0.094 \\
         & 28.407 -- 28.408 &   5 & 4.615 & 0.139 \\
         & 29.495 -- 29.497 &   5 & 4.909 & 0.403 \\
         & 30.488 -- 30.489 &   5 & 4.678 & 0.102 \\
         & 31.487 -- 31.488 &   5 & 4.580 & 0.098 \\
2000     &                  &     &       &       \\
January  &  3.475 --  3.477 &   5 & 4.625 & 0.123 \\
         &  8.458 --  8.459 &   5 & 4.730 & 0.309 \\
         & 10.449 -- 10.474 &   3 & 4.822 & 1.084 \\
         & 11.456 -- 11.458 &   5 & 4.757 & 0.140 \\
         & 14.379 -- 14.381 &   5 & 4.491 & 0.278 \\
         & 15.363 -- 15.371 &  22 & 4.716 & 0.768 \\
\hline
\end{tabular}
\end{center}
\end{table}

\begin{table}
\caption{Log of observations (5)}\label{tab:table5}
\begin{center}
\begin{tabular}{ccccc}
\hline
Month    & UT (start--end)  &  N  & mag   & error \\
\hline
2000     &                  &     &       &       \\
January  & 17.370 -- 17.374 &   5 & 3.386 & 0.475 \\
         & 18.370 -- 18.373 &   9 & 4.432 & 0.167 \\
         & 20.367 -- 20.371 &   9 & 4.164 & 0.393 \\
         & 21.378 -- 21.382 &  10 & 4.744 & 0.263 \\
         & 22.390 -- 22.391 &   3 & 3.988 & 0.477 \\
         & 24.375 -- 24.378 &   7 & 3.876 & 0.613 \\
         & 25.377 -- 25.382 &  14 & 4.641 & 0.265 \\
         & 26.376 -- 26.378 &   6 & 4.632 & 0.169 \\
         & 27.377 -- 27.379 &   5 & 4.565 & 0.205 \\
         & 28.393 -- 28.395 &   7 & 4.846 & 0.441 \\
         & 29.381 -- 29.384 &   7 & 4.675 & 0.192 \\
         & 30.383 -- 30.386 &   7 & 4.657 & 0.897 \\
         & 31.380 -- 31.382 &   7 & 4.934 & 0.505 \\
February &  1.395 --  1.397 &   6 & 4.984 & 0.427 \\
         &  2.390 --  2.392 &   5 & 4.453 & 0.156 \\
         &  4.393 --  4.393 &   2 & 4.905 & 0.088 \\
         &  5.387 --  5.388 &   5 & 4.703 & 0.090 \\
         &  7.385 --  7.386 &   5 & 4.647 & 0.263 \\
         &  8.388 --  8.390 &   2 & 5.152 & 0.256 \\
         &  9.396 --  9.398 &   5 & 4.826 & 0.216 \\
         & 10.391 -- 10.392 &   5 & 4.991 & 0.678 \\
         & 11.394 -- 11.395 &   3 & 3.995 & 0.396 \\
         & 12.390 -- 12.392 &   5 & 4.756 & 0.465 \\
         & 13.386 -- 13.387 &   5 & 4.090 & 0.665 \\
September& 18.732 -- 18.737 &  12 & 5.741 & 1.296 \\
         & 19.757 -- 19.762 &  12 & 5.642 & 0.134 \\
         & 20.752 -- 20.764 &  32 & 5.861 & 0.357 \\
         & 24.636 -- 24.639 &  10 & 5.667 & 0.168 \\
         & 26.785 -- 26.789 &  10 & 5.923 & 0.208 \\
         & 27.603 -- 27.606 &  10 & 5.939 & 0.177 \\
         & 28.619 -- 28.623 &  10 & 5.694 & 0.252 \\
         & 29.584 -- 29.587 &   9 & 5.084 & 0.132 \\
October  &  1.584 --  1.589 &  12 & 5.531 & 0.156 \\
         &  9.638 --  9.641 &  10 & 5.528 & 0.305 \\
         & 10.671 -- 10.676 &  12 & 5.682 & 0.328 \\
         & 11.617 -- 11.620 &  10 & 5.682 & 0.494 \\
         & 12.668 -- 12.671 &   3 & 4.665 & 0.854 \\
November & 22.552 -- 22.556 &  13 & 6.016 & 0.356 \\
December &  2.545 --  2.547 &   7 & 6.253 & 0.253 \\
\hline
\end{tabular}
\end{center}
\end{table}

\section{Results}
\begin{figure}
  \begin{center}
    \FigureFile(88mm,60mm){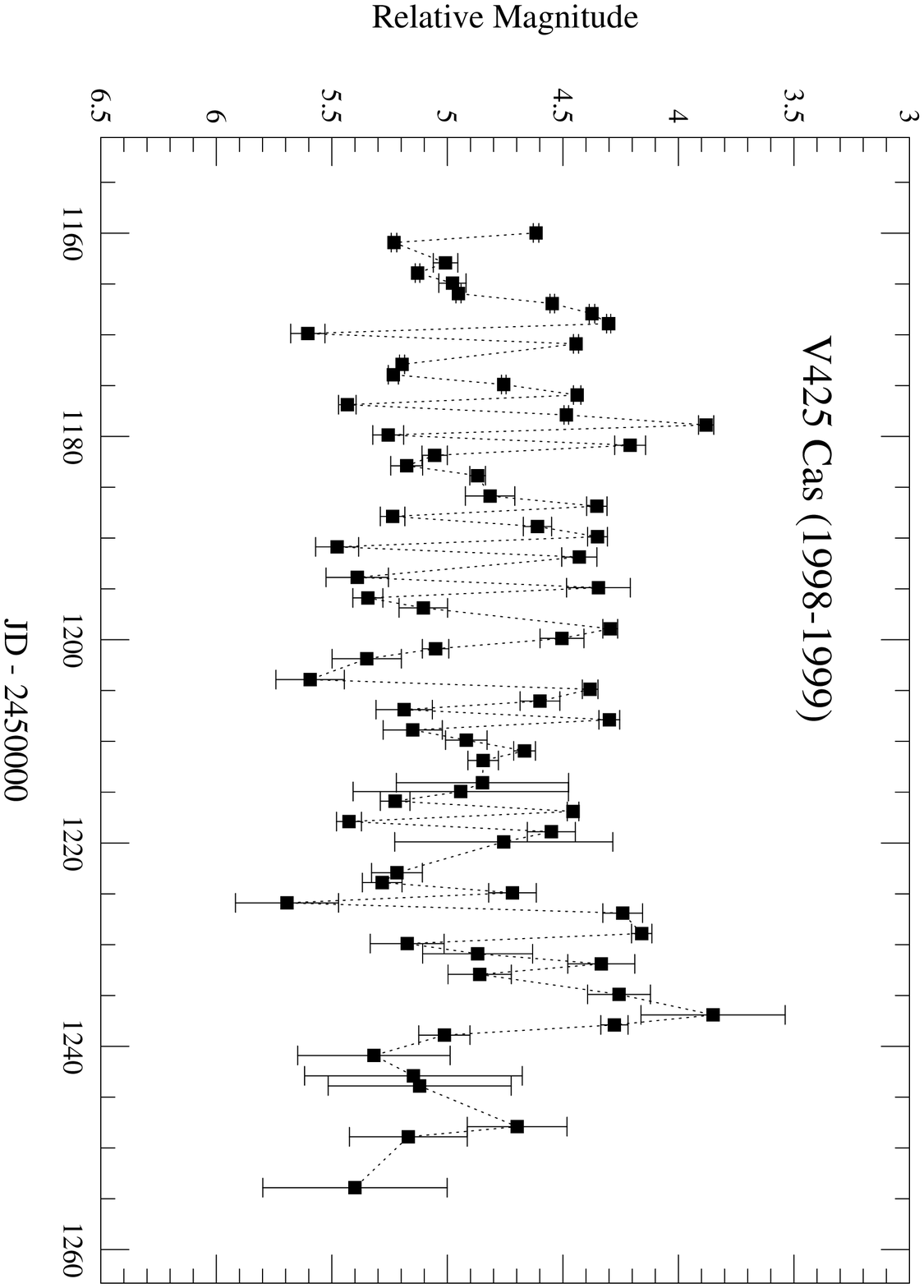}
  \end{center}
  \caption{Light curve of V425 Cas in the 1998--1999 season.  Each point
  and error bar represent nightly averaged magnitude and its standard
  error, respectively.  The magnitudes are given relative to GSC 3985.1444
  ($R_c=\sim11.0$).}
  \label{fig:figure1}
\end{figure}

  Fig. \ref{fig:figure1} shows the light curve of the 1998--1999 season,
when the object stayed 1.0--2.0 mag below the high-state level.
Large-amplitude, rapid variations are clearly seen with a time scale of
a few days.  Such a large-amplitude, short-term variation was never
observed in any hydrogen-rich CVs, including VY Scl-type stars.

\begin{figure}
  \begin{center}
    \FigureFile(88mm,60mm){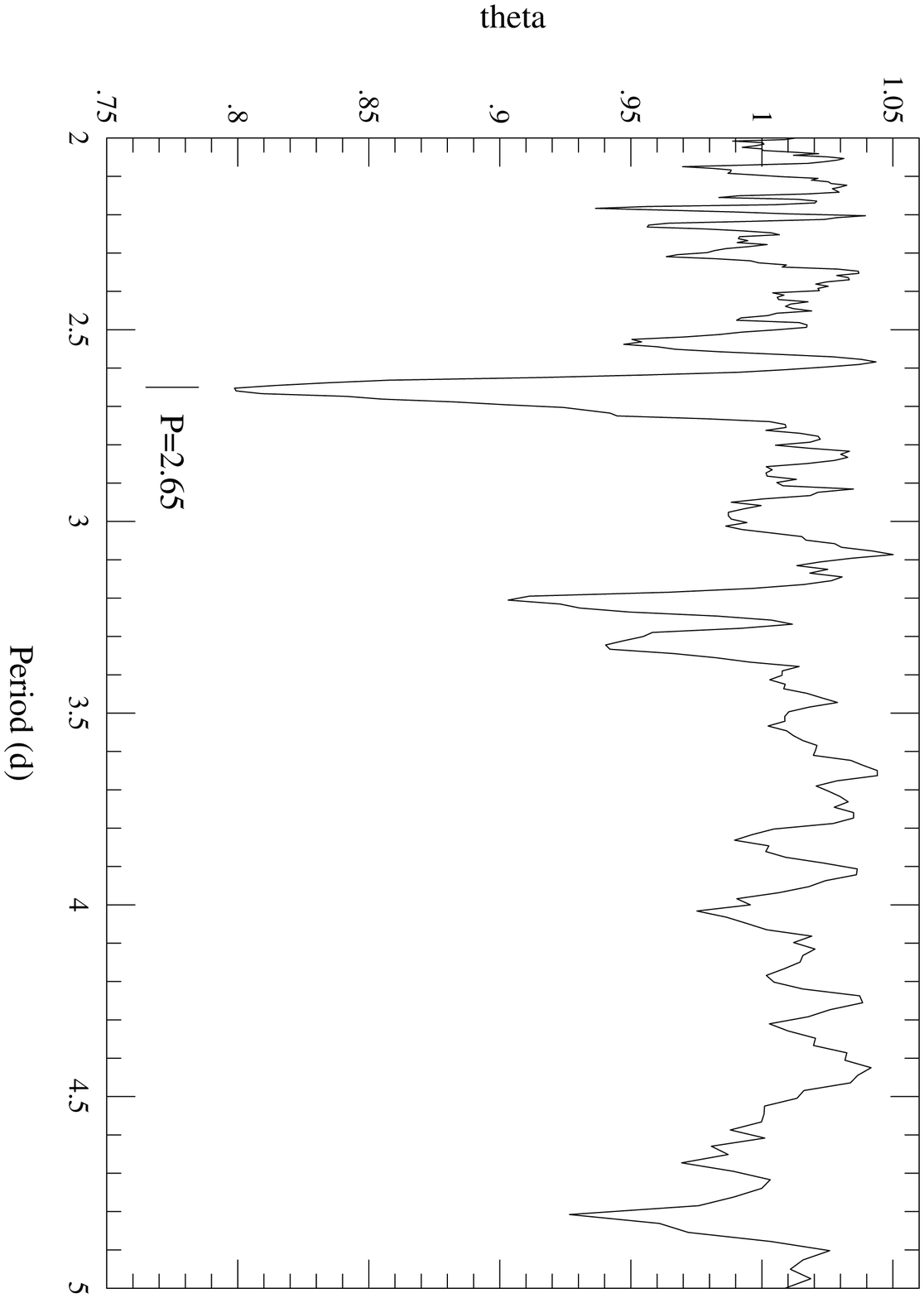}
  \end{center}
  \caption{Period analysis of V425 Cas in the 1998--1999 season.}
  \label{fig:figure2}
\end{figure}

  Fig. \ref{fig:figure2} shows the result of period analysis, using
the Phase Dispersion Minimization (PDM) method \citep{ste78}.
The strongest period is seen at $P=2.65$ d.  No other significant
period was found between $P=2$ d and $P=5$ d.  Since our observations
were basically sampled once per night, we can not completely rule out
the possibility of a shorter period.  We have searched signals around
the orbital period of V425 Cas (0.1496 d, \cite{sha83}) and found
no significant period representing the overall light variation.
This strongly indicates that the variation is not related to its
orbital motion.  The low inclination angle (25$^{\circ}$, \cite{sha83})
also makes unlikely orbital modulations as a major cause of variation.

\begin{figure}
  \begin{center}
    \FigureFile(88mm,60mm){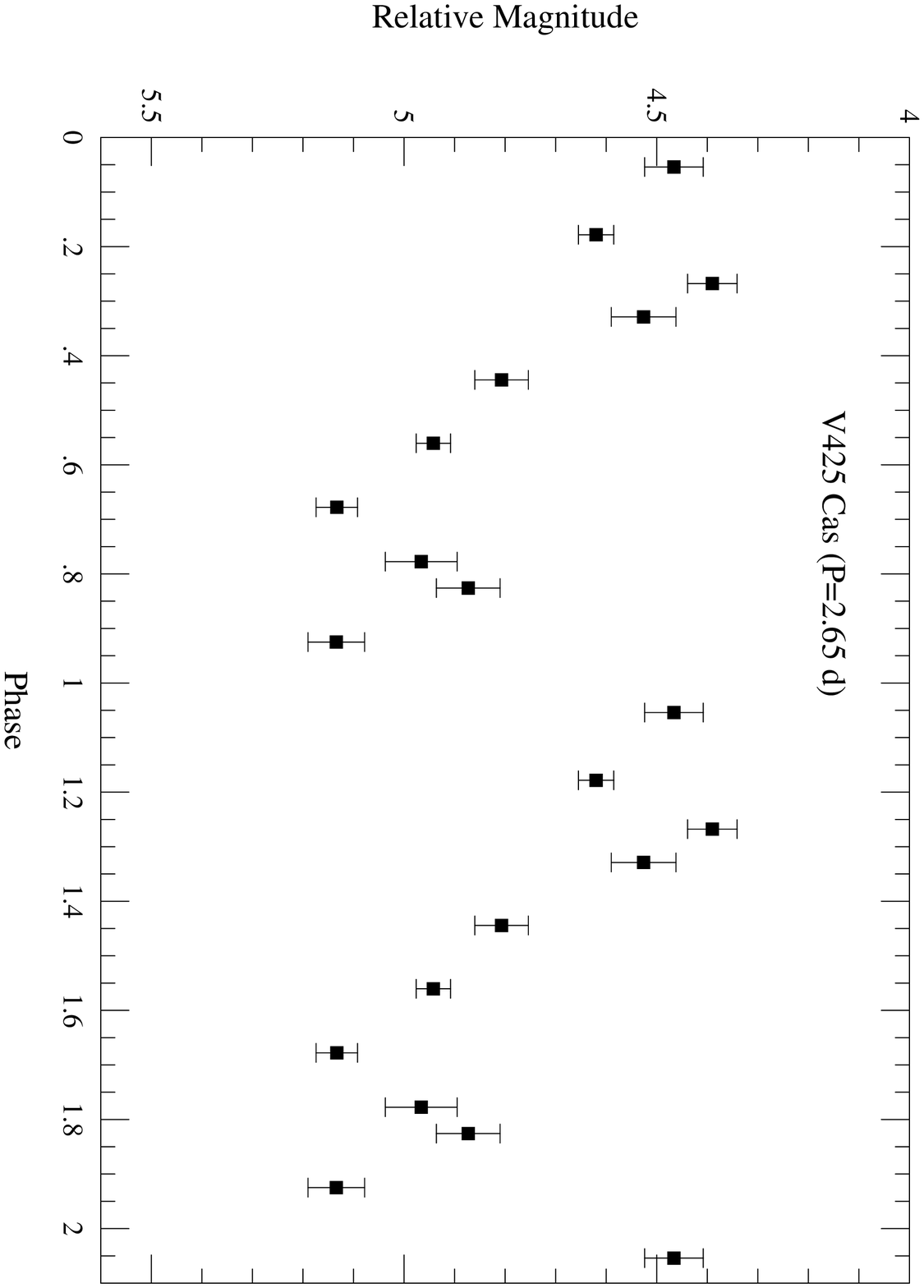}
  \end{center}
  \caption{Averaged light curve of V425 Cas folded by the period of 2.65 d
  in the 1998--1999 season.  The phase is taken arbitrarily.}
  \label{fig:figure3}
\end{figure}

  Fig. \ref{fig:figure3} shows the light curve in the 1998--1999 season
folded by this 2.65 d period.  The averaged light curve clearly shows
0.6--0.7 mag modulations, having a rapid rise and a slower decline.
The profile of the light curve resembles those of dwarf nova outbursts
with short recurrence times.

\begin{figure}
  \begin{center}
    \FigureFile(88mm,60mm){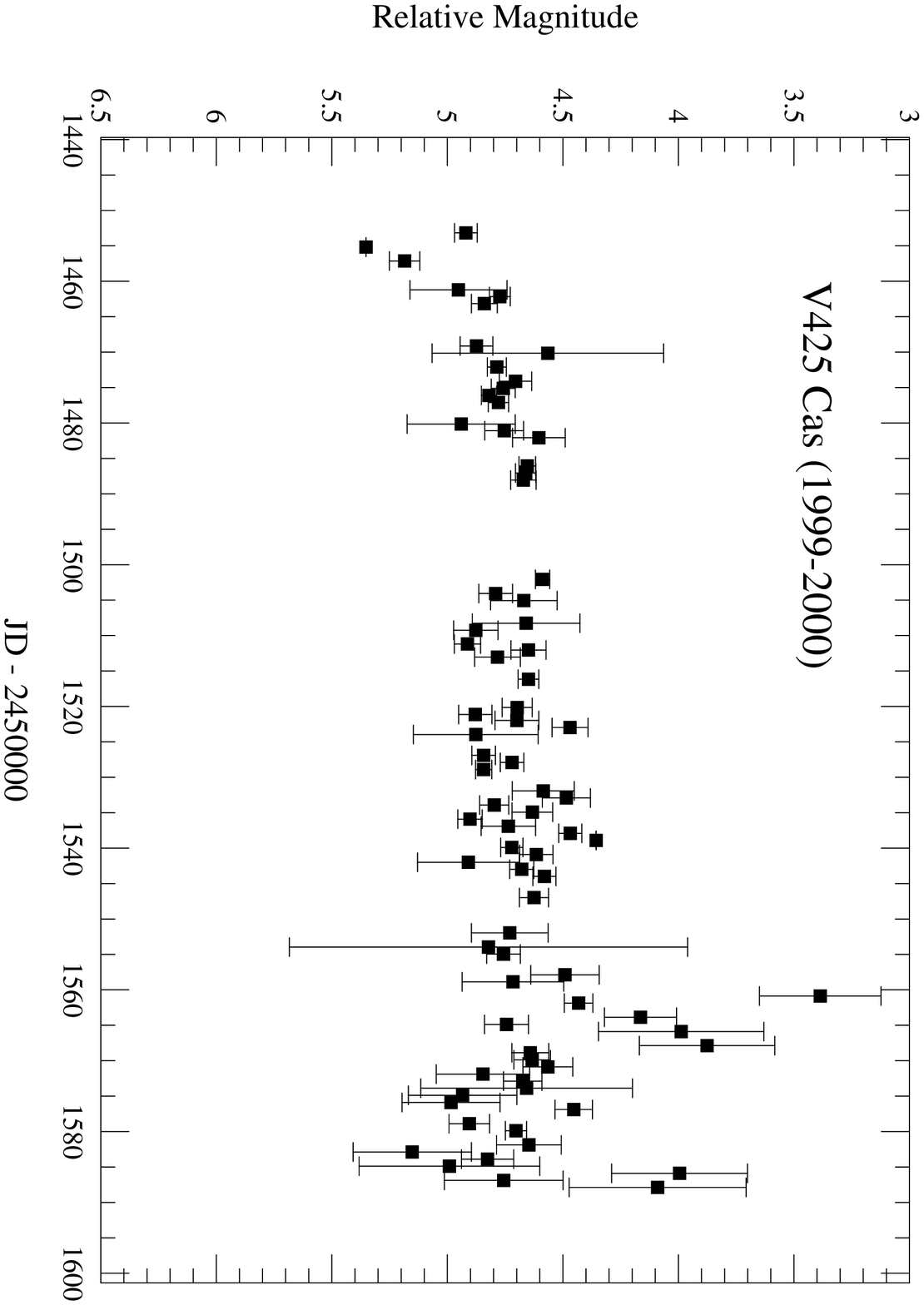}
  \end{center}
  \caption{Light curve of V425 Cas in the 1999--2000 season.  The vertical
  scale and the symbols are the same as in Fig. \ref{fig:figure1}.
  The 2.65-d oscillations completely disappeared.}
  \label{fig:figure4}
\end{figure}

  Fig. \ref{fig:figure4} shows the light curve in the 1999--2000 season,
in which the object was at a similar brightness level as in the 1998--1999
season (the averaged magnitudes of these two seasons agree within 0.05
mag).  Nevertheless, the 2.65-d period had completely disappeared.
Period analysis of the data did not yield any significant periodicity.

\begin{figure}
  \begin{center}
    \FigureFile(88mm,60mm){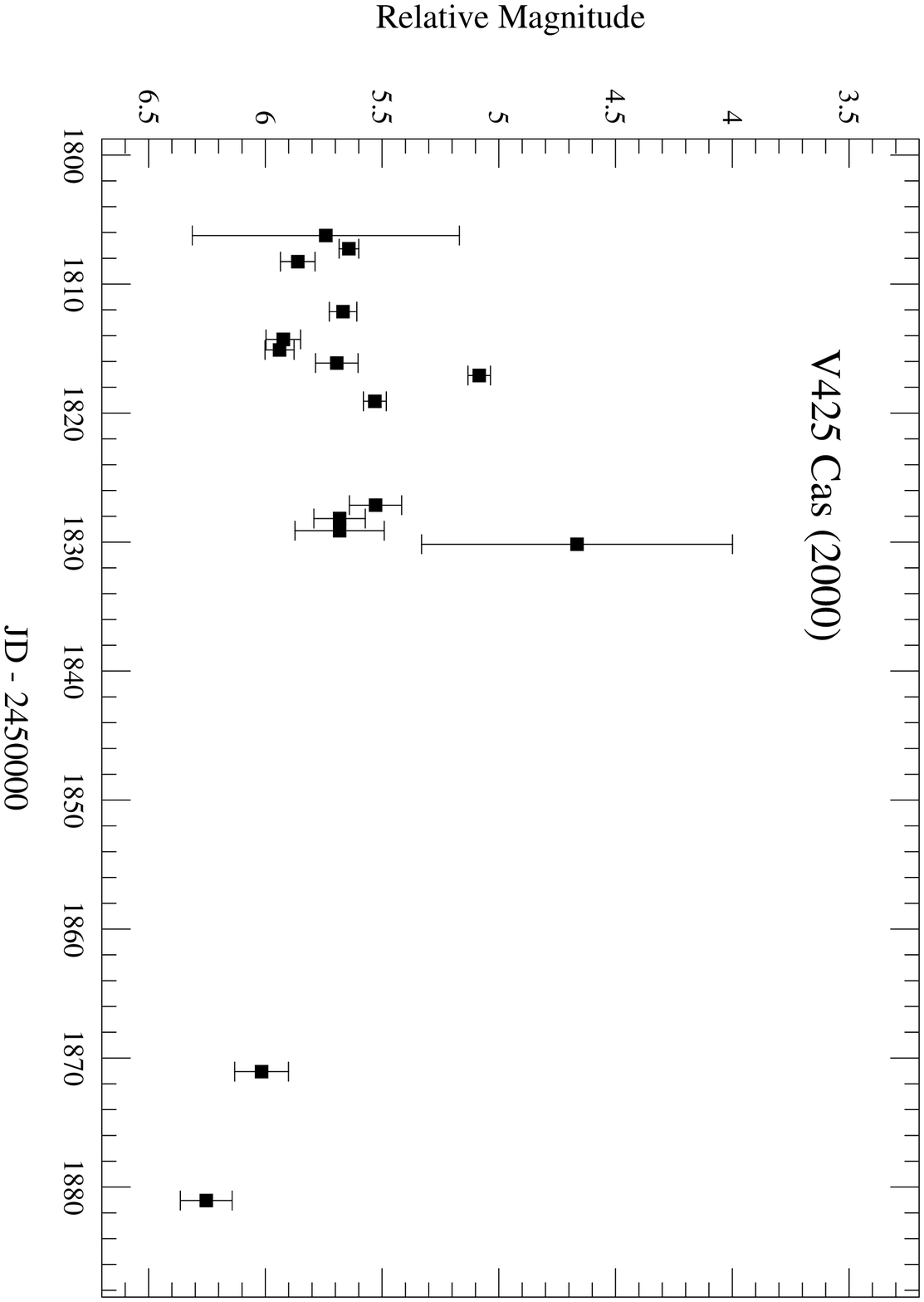}
  \end{center}
  \caption{Light curve of V425 Cas in the late 2000 season.  The object
  had faded considerably compared to the preceding two seasons.}
  \label{fig:figure5}
\end{figure}

  Fig. \ref{fig:figure5} shows the light curve in the late 2000 season.
The object had further faded by $\sim$1.0 mag at the beginning of this
season.  There is an evidence of a further fade in 2000 November -- December.
It is likely the object was entering a deep low state during this season.

\section{Discussion}
  The large amplitude (up to 1.5 mag, and 0.6--0.7 mag in average, which is
close to a factor of two flux variation) of the variations far exceeds those
of known (local) disk oscillations, such as quasi-periodic oscillations,
in CVs.  The amplitude more suggests that dwarf nova-type disk instability
is taking place, which is also consistent with the observed profile of
variation.  The main difference from the ordinary dwarf novae is the extreme
shortness of the recurrent time.  Dwarf novae with orbital periods around
that of V425 Cas have typical recurrence time of ten to a hundred of days,
the shortest known one being AM Cas having a recurrent time as short as 8 d
(cf. \cite{rit98}).
The recurrence time of dwarf novae is basically governed by the two
factors, namely the viscous drift time in low $\dot{M}$ systems and
the build-up time in the outer disk torus in high $\dot{M}$ systems
(cf. \cite{osa96}).  The shortest limit of recurrence time is regulated
by the latter factor, which makes the recurrence time approximately
inversely proportional to $\dot{M}$.  Above the critical $\dot{M}$, the
disk becomes thermally stable, and this determines the shortest limit.
According to calculations by \citet{ich94}, the expected minimum period
is slightly below $\sim$10 d, which is in good agreement with the
AM Cas case.  A different mechanism is therefore needed to explain the
extremely short recurrent time in V425 Cas.

  A clue to this problem can be found in another example of a striking
departure of recurrence time from the disk-instability theory: ``purr"
type outbursts observed during the outburst of an X-ray transient,
V518 Per = GRO J0422+32 \citep{min94}, which had a recurrence time of
$\sim$10 d, which is several times to an order of magnitude shorter than
the expected recurrence time from the disk instability model.  \citet{min94}
argued that X-ray irradiation on the accretion disk can effectively
increase the disk temperature, and suppress the instability in inner
portions of the disk, producing small-scale purr-type outbursts
(for model calculations, see \cite{min90}).  This scheme could apply to
accretion disks in CVs, if there is an appropriate source of irradiation.
\citet{min94} required an X-ray luminosity of $L_{\rm X}=10^{35}$
erg s$^{-1}$ in producing purr-type outbursts.  Since the observed X-ray
luminosity during a low state of the VY Scl-type star V751 Cyg was
estimated to be $L_{\rm X}=10^{34-36}$ erg s$^{-1}$ \citep{gre99},
the effect of irradiation in VY Scl-type systems is expected to be enough
to produce similar purr-type outbursts in CVs (CVs generally have smaller
accretion disks than in X-ray transients, which would make even a lower
$L_{\rm X}$ to work equally efficiently).  We thus regard the present
discovery of dwarf nova-like oscillations in V425 Cas as another promising
evidence for the effect of irradiation in VY Scl-type systems.
The disappearance of this kind of oscillations in the 1999--2000 season
may be explained, in the same context, by the reduction of irradiation,
due to the exhaustion of the accreted matter after a long-lasting of
low-accretion state.  The termination of accretion or the exhaustion of
the accreted matter probably led to a further fade observed within
the following eight months.

   Yet another intriguing similarity is found in dwarf nova-like
oscillations in helium CVs (AM CVn stars).  Two examples are known:
CR Boo ($P\sim$19 hr, \cite{pat97}) and V803 Cen ($P=$20--23 hr,
\cite{pat00}, \cite{kat01}).  \citet{pat97} proposed that this oscillation
in CR Boo can be understood as an extension of outbursts in hydrogen-rich
dwarf novae, using the known relations (Kukarkin-Parenago's relation
and Bailey's relation) in hydrogen-rich dwarf novae.
However, the assessment of the period should require calculations
in helium accretion disks.  \citet{tsu97} applied the dwarf nova-type
thermal and tidal instability model to the helium disk systems, and
obtained a recurrence time of $\sim$4--8 d, which is too long to explain
the observed oscillations in CR Boo and V803 Cen.  This recurrence
time was more appropriately demonstrated in ``normal outbursts" of
CR Boo \citep{kat00} between its superoutbursts recurring with a period
of 46 d.  \citet{kat00} suggested that lower-amplitude, short-period
oscillations observed by \citet{pat97} may result from the reflection of
the cooling and heating waves in the outer region of the accretion disk,
and not the full-disk outburst.  The mechanism responsible for this
effect was not clear at the time of the suggestion by \citet{kat00},
but the relatively strong X-ray flux in both stars inferred from ROSAT
observations, the effect of X-ray irradiation would be a promising
explanation.  Emerging evidences on the observable effect of irradiation
in CVs on their outburst behavior, as demonstrated in the present
observation, may be also a key to understanding the mystery of outbursts
in helium CVs.

\section{Conclusion}
   Our long-term CCD photometry of V425 Cas, during its faint state
in 1998--2000, revealed the presence of totally unexpected, unprecedented
large-amplitude (0.5--1.5 mag) oscillations with a time scale of
a few days.  The period analysis of oscillations has demonstrated that
this oscillation is well represented by a single period of 2.65 d.
The amplitude and mean profile of variation strongly suggests that
variation is caused by dwarf nova-type disk instability.  The shortness
of the period is more difficult to explain.  By taking recent discoveries
of supersoft X-rays in a low state VY Scl-type star into account,
we proposed that the suppression of the thermal instability of the inner
accretion disk by irradiation can be responsible for this large-amplitude
variation in V425 Cas.  This effect of irradiation may also apply to
helium cataclysmic variables, which show similar short-period,
large-amplitude oscillations, which were proposed to be dwarf nova-type
outbursts.

Part of this work is supported by a Research Fellowship of the
Japan Society for the Promotion of Science for Young Scientists (MU).

\end{document}